# Size-based ion selectivity of micropore electric double layers in capacitive deionization electrodes

Matthew E. Suss[1]

1. Faculty of Mechanical Engineering, Technion – Israel Institute of Technology, Haifa, Israel

**Abstract**

Capacitive deionization (CDI) is a fast-emerging technology most commonly applied to brackish water desalination. In CDI, salt ions are removed from the feedwater and stored in electric double layers (EDLs) within micropores of electrically charged porous carbon electrodes. Recent experiments have demonstrated that CDI electrodes exhibit selective ion removal based on ion size, with the smaller ion being preferentially removed in the case of equal-valence ions. However, state-of-the-art CDI theory does not capture this observed selectivity, as it assumes volume-less point ions in the micropore EDLs. We here present a theory which includes multiple counterionic species, and relaxes the point ion assumption by incorporating ion volume exclusion interactions into a description of the micropore EDLs. The developed model is a coupled set of nonlinear algebraic equations which can be solved for micropore ion concentrations and electrode Donnan potential at cell equilibrium. We demonstrate that this model captures key features of the experimentally observed size-based ion selectivity of CDI electrodes.

**Introduction**

Capacitive deionization (CDI) is an emerging technology most commonly applied to brackish water desalination, but also used towards other applications such as water softening, wastewater remediation, microfluidic sample preparation, and organic stream remediation.[1–4] Typical CDI cells employ two microporous carbon electrodes, and a separator layer between the electrodes which also serves as the feedstream flow channel (Fig. 1).[5] When the electrodes are charged via application of a cell voltage or current, electric double layers (EDLs) in the micropores electrosorb counterions, which results in the desalination of the feedstream. Compared to more established desalination technologies, such as reverse osmosis (RO) and flash distillation (FD), CDI does not require high pressure pumps or heat sources, and thus CDI systems can be highly scalable and energy efficient.[6] In addition to the typical CDI cell architecture, variations in cell design and materials such as flow-through electrodes CDI,[7,8] membrane CDI,[9,10] flow electrode CDI,[11,12] fluidized bed CDI,[13,14] hybrid CDI,[15] inverted CDI,[16,17] induced-charge CDI,[18] and CDI with intercalation electrodes[19,20] enable novel functionalities and performance enhancements. Micropores in CDI electrodes represent a highly confined geometry, where typically the pore size is on the order of the hydrated ion size. This confinement allows for optimization of the electrode's maximum salt adsorption capacity (mSAC) by maximizing the salt ion concentration in the micropore volume.[21] Thus, in the micropore, the interplay between ion size and shape, ion hydration, and pore size and shape is of high importance towards predicting and optimizing electrode storage capacity, and predicting the concentrations of ionic species in the charged micropore EDL.

A promising feature of CDI electrodes is their demonstrated ability to preferentially electrosorb specific counterions from a feedstream with multiple counterionic species. For example, CDI electrodes have demonstrated selective electrosorption based on ion valence,[22] size,[23–25] and shape.[26] Selectivity is a desirable feature in water treatment, in particular for removing ionic contaminants from ground or wastewaters. Certain ions, such as fluorine ($F^-$), nitrate ($NO_3^-$), and heavy metal-type ions such as ferric ($Fe^{3+}$) and chromate ($CrO_4^{2-}$), can be dangerous to human health even at low concentrations, and thus their selective removal enables energy efficient treatment of affected waters.[24,27–29] Beyond water purification, ion selectivity in CDI cells can in the future be leveraged towards the development of a versatile separations technology, with applications beyond simply separating salt from water.[2] Thus, theoretical models are required to understand the mechanisms underpinning the observed selectivity, and to predict and optimize selectivity in feedstreams containing many ions which require various levels of removal. Current CDI models predict a micropore EDL selectivity based on ion valence, for example, finding that at cell equilibrium divalent ions can be selectively stored over monovalent ions for the case of divalent calcium ions ($Ca^{2+}$) and monovalent sodium ions ($Na^+$).[22] Significant experimental evidence has demonstrated that CDI systems can also selectively remove smaller univalent ions at the expense of larger univalent ions, such as the selective removal of smaller potassium ion ($K^+$) compared to larger $Na^+$.[23,24,30] However, state-of-the-art CDI theory considering multiple counterionic species assumes ions stored in the micropore volume are point (volume-less) ions, and thus these models cannot predict selectivity based on ion size (Fig. 1a).[31,32] This work aims to rectify this mismatch between experimental measurements and model predictions, by updating CDI theory for systems with multiple counterions to include ion volume exclusion interactions in the description of micropore EDLs (Fig 1b). We then show that our theoretical predictions of selectivity in systems with two counterions of differing size captures the essential trends observed experimentally. Finally, our model predicts that CDI electrodes can be strongly selective even at modest size ratios between counterions, which has implications for modeling pH distributions in CDI systems and towards the utility of CDI cells as a versatile ionic separations platform.

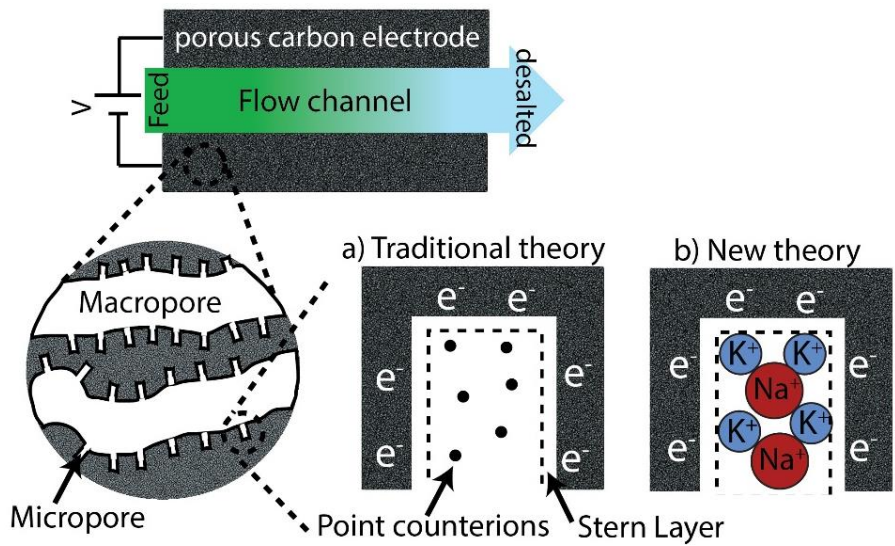

**Figure 1:** Schematic of a capacitive deionization (CDI) cell, with insets depicting the electrode's pore structure consisting of macropores and micropores, and charged micropore electric double layers (EDLs). Inset a) depicts a charged micropore for the case of volume-less point ions, an assumption of traditional CDI theory. Inset b) shows a charged micropore with ions instead treated as hard spheres with varying radii, which allows for an exploration of ion size effects on ion selectivity and desalination by CDI.

**Theory**

We begin with the theory for equilibrium salt electrosorption by micropore EDLs assuming point (volume-less) ions (Fig. 1a), as is typically assumed in CDI theory with multiple counterionic species.[31] Invoking equilibrium, and so equalized electrochemical potential, between the micro and macropores results in:

$$\ln c_{mi,i} / c_{ma,i} + z_i \Delta \phi_D = 0 . \quad [1]$$

Here $c_{mi,i}$ and $c_{ma,i}$ are the concentrations of ion $i$ in the micropore and macropore, respectively, $\Delta \phi_D$ is the dimensionless Donnan electric potential defined as $\phi_{mi} - \phi_{ma}$, and $z_i$ is the valence of ion $i$. All electric potentials are non-dimensionalized by the thermal potential $V_T = kT/e$, where $T$ is the temperature, $k$ is the Boltzmann constant, and $e$ is the electron charge. We now restrict the discussion to an electrolyte consisting of two counterions denoted by subscript "1" and subscript "2". To focus on the effect of ion size on selectivity, we assume both counterions have equal valence, $z$. For simplicity, we assume the electrolyte consists of a single coion with valence $-z$, denoted by subscript "co". Applying Eq. 1 to each ion, and invoking electroneutrality in the macropore, we obtain the following:

$$c_{mi,1} / c_{ma,1} \equiv \alpha_1 = \exp(-z \Delta \phi_D) , \quad [2]$$

$$c_{mi,2} / c_{ma,2} \equiv \alpha_2 = \exp(-z \Delta \phi_D) , \quad [3]$$

$$c_{mi,co} = (c_{ma,1} + c_{ma,2}) \exp(z \Delta \phi_D) . \quad [4]$$

The parameter $\alpha_i$ is defined as the ratio of the concentration of counterion $i$ in the charged micropore EDL to its concentration in the macropore. The ratio $\alpha_1 / \alpha_2$ has previously been termed a selectivity ratio,[33] and will be used here to describe the selectivity of the micropore EDL at a given charge state (given Donnan potential). We can see from Eqs. 2 and 3 that $\alpha_1 / \alpha_2 = 1$ for all values of $\Delta \phi_D$, indicating that the micropore EDL does not preferentially select either counterion for the case of equal valence and point counterions. Other works have instead used ratios of excess counterion concentrations to quantify selectivity, which accounts for both electrostatic and entropic effects (e.g. the selective storage of a counterion due to a higher macropore concentration relative to the other counterion),[34,35] while the selectivity ratio used in this work accounts for solely the electrostatic effect.

By combining Eqs. 2-4, we can express the micropore volumetric charge density as:[31]

$$\sigma_{mi} \equiv \sum_i z_i c_{mi,i} = -2z(c_{ma,1} + c_{ma,2}) \sinh(z \Delta \phi_D) . \quad [5]$$

Alternatively, by the definition of the Stern layer capacitance, micropore volumetric charge density can be expressed as:[31]

$$\sigma_{mi} = -\frac{C_{st} \Delta \phi_{st} V_T}{F} . \quad [6]$$

Here $C_{st}$ is the volumetric stern layer capacitance, and $\Delta \phi_{st}$ is the dimensionless potential drop across the Stern layer, and the negative sign is inserted as the micropore charge is opposite in sign to the wall charge. Setting

Eqs. 5 and 6 equal to each other, using the macropore potential as a reference ($\phi_{ma} \equiv 0$), and using $\phi_e = \Delta\phi_{st} + \Delta\phi_D$ where $\phi_e$ is the electrode's dimensionless solid phase potential,[31] we obtain:

$$\Delta\phi_D = \phi_e - \frac{2zF\sinh(z\Delta\phi_D)}{C_{st}V_T}(c_{ma,1} + c_{ma,2}) \,. \tag{7}$$

Eq. 7 is can be solved for $\Delta\phi_D$, after which all the micropore concentrations given by Eqs. 2-4, and the selectivity ratio can be determined. Eq. 7 can be linearized for the case where $z\Delta\phi_D \ll 1$, however, such a low Donnan potential is atypical for CDI electrodes (see Fig 3c).

While Eqs. 2, 3, and 7 can be used to solve for the selectivity ratio and micropore concentrations for the case of point ions (Fig. 1a), we now develop an analogous set of equations which consider finite-sized ions by accounting for ion volume exclusion interactions in the micropore EDL (Fig. 1b). To account for volume exclusion effects, we add a dimensionless excess chemical potential difference between the micro and macropores, $\Delta\mu_i^{ex}$, to the equilibrium expression given by Eq. 1:[36]

$$\ln(c_{mi,i}/c_{ma,i}) + z_i\Delta\phi_D + \Delta\mu_i^{ex} = 0 \,. \tag{8}$$

For a system with two equal charge and finite-sized counterions, applying Eq. 8 we obtain:

$$c_{mi,1}/c_{ma,1} \equiv \alpha_1 = \exp(-z\Delta\phi_D - \Delta\mu_1^{ex}) \,, \tag{9}$$

$$c_{mi,2}/c_{ma,2} \equiv \alpha_2 = \exp(-z\Delta\phi_D - \Delta\mu_2^{ex}) \,, \tag{10}$$

$$c_{mi,co} = (c_{ma,1} + c_{ma,2})\exp(z\Delta\phi_D - \Delta\mu_{co}^{ex}) \,. \tag{11}$$

From Eqs. 9 and 10, the selectivity ratio can be simplified to $\alpha_1/\alpha_2 = e^{(\Delta\mu_2^{ex} - \Delta\mu_1^{ex})}$. Thus, the selectivity ratio is no longer necessarily unity, and the micropore EDL will preferentially select the ion with the smaller value of excess potential. To account for ion volume exclusion interactions, we use an appropriate analytical expression for $\mu_i^{ex}$ which can be applied to ion $i$ in the macropore or micropore. Two such equations have been extensively employed for planar EDLs and electrolytes with multiple counterions, the Bikerman equation derived from lattice-gas type model for the EDL, and the Boublik-Mansoori-Carnahan-Starling-Leland (BMCSL) equation derived by considering ions as hard spheres with differing diameters (an extension of the Carnahan-Starling equation of state for single-sized hard spheres).[34,35,37] While the Bikerman equation is the simpler expression, it has been well-established that the BMCSL equation (and the related Carnahan-Starling equation) is the more accurate prediction for the case of EDLs with planar geometry.[35,38] Thus, we here employ the BMCSL equation, to our knowledge for the first time, to study size-based selectivity by micropore EDLs of CDI electrodes. The BMCSL equation is given by:

$$\mu_i^{ex} = -\left(1 + \frac{2\zeta_2^3 d_i^3}{\phi^3} - \frac{3\zeta_2^2 d_i^2}{\phi^2}\right)\ln(1-\phi) + \frac{3\zeta_2 d_i + 3\zeta_1 d_i^2 + \zeta_0 d_i^3}{1-\phi} + \frac{3\zeta_2^2 d_i^2}{(1-\phi)^2}\left(\frac{\zeta_2}{\phi} + \zeta_1 d_i\right)$$
$$- \zeta_2^3 d_i^3 \frac{\phi^2 - 5\phi + 2}{\phi^2(1-\phi)^3} \,. \tag{12}$$

Here $d_i$ is the hard-sphere diameter of ion $i$, which we consider to be an adjustable parameter used to fit data, as with previous works employing the BMCSL equation.[35] Further, $\phi = \sum_i \phi_i = \sum_i \left(\pi d_i^3/6\right) c_i N_a$ represents the volume fraction occupied by all finite sized ions, $N_a$ is Avogadro's constant, and $\zeta_k = \sum_i \phi_i d_i^{k-3}$. We now invoke the definitions of volumetric micropore charge density to obtain:

$$\Delta\phi_D = \phi_e + \frac{zF}{C_{st}V_T}\left[c_{ma,1}(e^{-z\Delta\phi_D - \Delta\mu_1^{ex}} - e^{z\Delta\phi_D}) + c_{ma,2}(e^{-z\Delta\phi_D - \Delta\mu_2^{ex}} - e^{z\Delta\phi_D})\right]. \quad [13]$$

Eqs. 9, 10, 12 and 13 form a set of algebraic equations which must be solved together for unknowns $\Delta\phi_D$, $c_{mi,1}$, $c_{mi,2}$. The model presented here can be used to predict the salt stored in an electrode's micropore EDLs at cell equilibrium (i.e. at steady state after a constant voltage cell charging has completed). To predict dynamic CDI data, such as cell effluent composition versus time after application of a cell voltage or current, the model presented here must be generalized to include two electrodes and coupled to a set of macroscopic transport equations.[5] We will leave the latter developments to a future work, while here focusing on the selectivity predicted at cell equilibrium.

**Results**

While the selectivity ratio, $\alpha_1/\alpha_2$, is convenient to study the theoretically expected selectivity, a related metric known as the separation factor is more readily measured experimentally. We denote the separation factor by $\beta_1/\beta_2$, where $\beta_i$ is obtained from experimental data via the expression $\beta_i = SAC_i/c_{feed,i}$, and $SAC_i$ is the cells' salt adsorption capacity for ion $i$ in units of mol/g$_{carbon}$.[39] $SAC_i$ is obtained from measurements of the concentration of ion $i$ in the cell effluent during the charge step in a single-pass experiment, which when subtracted from the feed concentration, integrated in time, and multiplied by feed flow rate gives the total moles of ion $i$ removed from the feed.[1] The charge step when measuring $SAC_i$ generally begins with an uncharged electrode and ends at cell equilibrium. To calculate the selectivity ratio from our model results, we use the expression $\beta_1/\beta_2 = \left[(c_{mi,1} - c_{ma,1})*c_{ma,2}\right]/\left[(c_{mi,2} - c_{ma,2})*c_{ma,1}\right]$, which assumes $c_{ma,i} = c_{feed,i}$ at the beginning and end of the charge step, and the initial (pre-charging) micropore concentration equals the macropore concentration (i.e. no micropore chemical charge[17]). For the typical CDI operating condition of high Donnan potentials, $z\Delta\phi_D \gg 1$, our model predicts that $\beta_1/\beta_2 \sim \alpha_1/\alpha_2$ since $c_{mi,i} \gg c_{ma,i}$ (for e.g. see Fig. 3a).

In Figure 2, we compare model predictions for $\beta_1/\beta_2$ (dashed line and open markers), to the measured separation factor previously observed at cell equilibrium for three experimental CDI cells (filled markers).[23,30,32] The experiments considered all utilized electrolytes containing two competing, univalent ions, either Cl$^-$/F$^-$ or K$^+$/Na$^+$. For $\beta_1/\beta_2$, we define counterion "1" to be the smaller ion and counterion "2" as the larger ion based on the known hydrated ion radius in bulk electrolyte, from ref. [40]. For each experiment, the observed separation factor is significantly above unity, ranging between 1.22 and 1.28, indicating that the smaller of the two competing univalent ions is preferentially electrosorbed by the micropore EDLs. The predicted separation factor by traditional CDI theory assuming point ions is unity (dashed line in Fig. 2), and thus cannot capture this experimentally-observed selectivity. Model predictions including ion volume exclusion effects fall within the range of 1.22 to 1.3, when assuming a typical Stern layer capacitance of 0.2 GF/m$^3$,[23,24,41] and the counterion hard-sphere ion diameter $d_i$ as 1.25 times the hydrated ion diameter in bulk solution.[40] If we had instead used simply the hydrated ion diameter in bulk solution as $d_i$, the predicted separation factor would have been between 1.09 and 1.12 (not shown in Fig. 2). The use of larger hard-sphere diameter than hydrated ion diameter has been also required to fit BMCSL-based theory to data for the case of a planar EDL.[34,35] For example, Biesheuvel et al. required a hard-sphere diameter of 1.15-1.2 times the hydrated ion diameter in bulk solution in order to allow for theory predictions to approach experimentally-observed selectivity.[35] Thus, our results are in-line with the previously-stated hypothesis that the effective hydrated counterion size near highly-charged interfaces may be significantly larger than in bulk solution due to increased electrostatic repulsion between counterions within such EDLs.[34] For simplicity, in these calculations we neglected the (small) excess potential acting on the coions, as predicted coion concentration in micropore EDLs at high electrode potential approaches zero. For the electrode potential used in the theory calculations, we assumed that the experimental cells were symmetric so that $\phi_e = V_{cell}/2V_T$ at cell equilibrium, where $V_{cell}$ is the applied cell voltage during experiments. This latter assumption is reinforced by ample experimental evidence that CDI cells with the same, chemically uncharged, microporous carbon material as anode and cathode typically exhibit an approximately symmetric voltage distribution, despite the slight size variations between the anion and cation (typically sodium and chloride).[19,42,43] Overall, both the model and experimental results of Fig. 2 demonstrate that ion size plays a significant role in the CDI process for electrolytes with competing, equally-charged ions, even when there are only slight differences in ion size (e.g. a <0.4 Å difference for Na$^+$ and K$^+$).

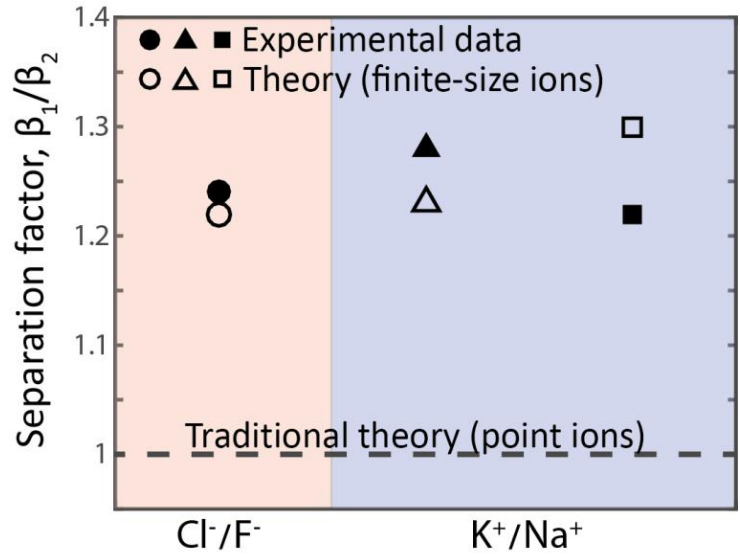

**Figure 2**: Plot of separation factor, $\beta_1 / \beta_2$ observed experimentally in literature for electrolytes with either Cl$^-$/F$^-$ (circle marker/Tang et al.[32]) or K$^+$/Na$^+$ (triangle marker/Dykstra et al.[23], and square marker/Hou et al.[30]) as competing ions. The dashed line represents the results of traditional CDI theory assuming point ions, where

$\beta_1 / \beta_2 = 1$ in all cases considered. When accounting for ion volume exclusion effects via the BMCSL equation (open markers), our model predicts a greater-than-unity separation factor as is also observed experimentally.

In Figure 3, we show model predictions for the case of Na⁺ and K⁺ counterions, a Stern capacitance of 0.2 GF/m³, macropore concentration of 10 mM for each counterion, and the same hard-sphere diameters used in Fig. 2 for Na⁺ and K⁺. Figure 3a shows the predictions of micropore concentration versus non-dimensional electrode potential for the case of point ions (dashed line) and finite-size ions (red and blue lines). For the case of point ions, the predicted concentration profiles for Na⁺ and K⁺ are identical, as expected from Eqs. 2 and 3. But, when accounting for volume exclusion interactions, we see significant differences in micropore concentration at typical CDI electrode potentials, despite the small difference in size between sodium ($d_{Na^+} = 0.9\,nm$) and potassium ($d_{K^+} = 0.83\,nm$).[40] For example, at $\phi_e = -25$, the predicted micropore concentration of Na⁺ is about 410 mM, while that of K⁺ is 565 mM. We do not show on the plot predicted coion concentration, as this concentration is below 10 mM for a charged electrode and approaches zero at high voltages. In Figure 3b, we show predictions of micropore charge density versus electrode potential, which shows non-negligible deviations in micropore charge between the case of point ions and finite-size ions at high electrode potentials. The case of finite-sized ions achieves lower micropore charge for a given voltage, which demonstrates that volume exclusion interactions act to lower the capacitance of the electrode. We can quantify the relative importance of volume exclusion effects by comparing these to entropic and electrostatic contributions to the electrochemical potential. In Figure 3c, we plot the predictions for the net components of electrochemical potential acting on the finite-sized K⁺ cation, including the net entropic component, $\ln\left(c_{mi,K^+} / c_{ma,K^+}\right)$, the net electrostatic component, $z\Delta\phi_D$, and net volume exclusion effects, $\Delta\mu_{K^+}^{ex}$. As expected, the electrostatic component drives ion storage in the micropore as this term is negative for all electrode potentials. Due to the equilibrium between micro and macropore, the electrostatic contribution is balanced by both entropic and volume interaction effects. Interestingly, at high electrode potentials volume effects are significant relative to entropic, even for the relatively small-sized potassium ion. For example, at $\phi_e = -25$, the entropic component contributes about 4 kT per ion, while volume effects amount to over 2 kT per ion. The latter result reinforces that accounting for ion volume effects in micropores is likely crucial towards accurate modeling of desalination and electrode capacitance in CDI systems.

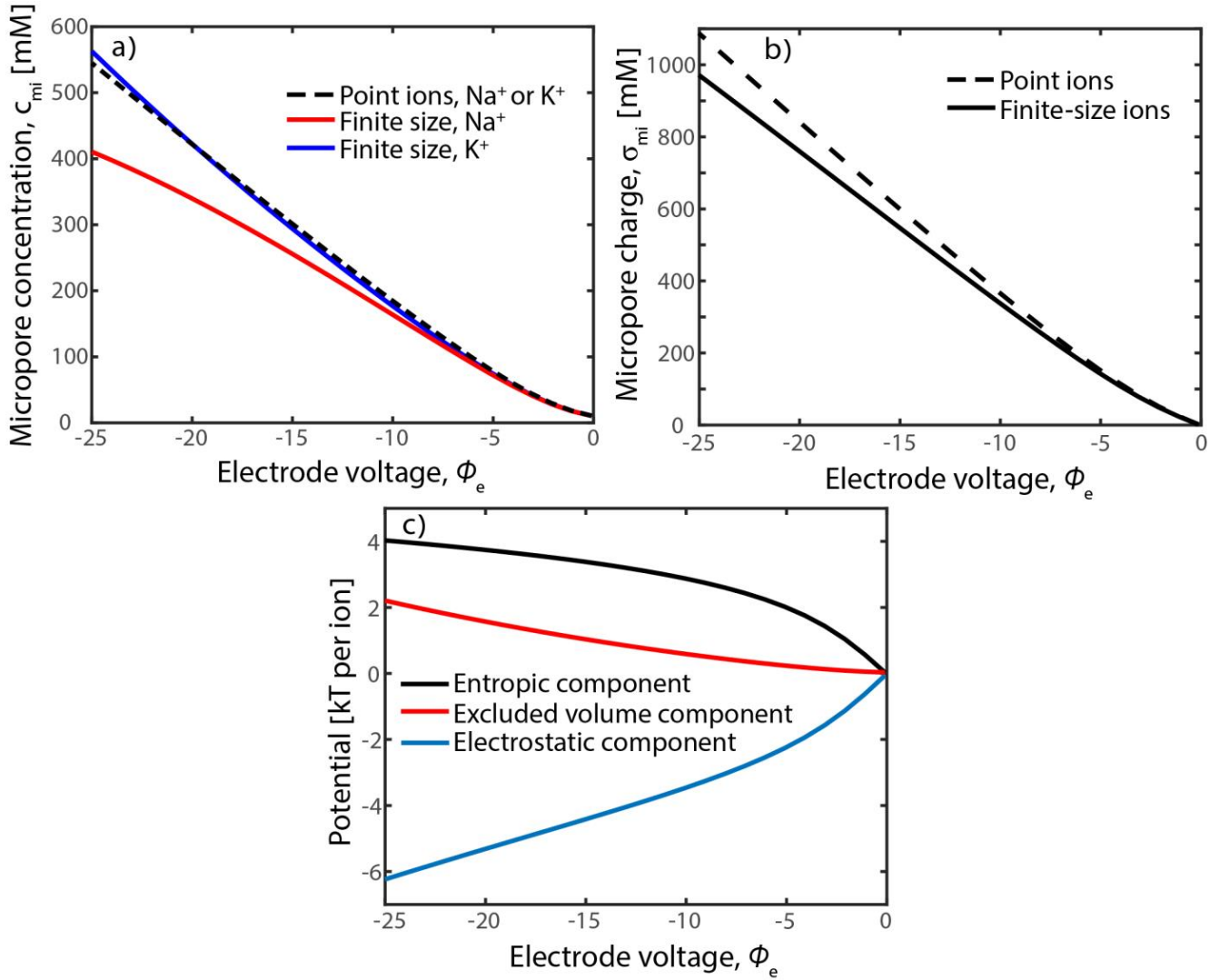

**Figure 3**: a) Predictions of micropore counterion concentrations for an electrolyte with Na$^+$ and K$^+$ as counterions versus dimensionless electrode potential $\phi_e$. Model parameters include $c_{ma}$ = 10 mM for each counterion, $C_{st}$ = 0.2 GF/m$^3$, and hard-sphere diameters $d_{Na^+} = 0.9\ nm$ and $d_{K^+} = 0.83\ nm$. The plot shows the results of traditional theory assuming point ions (dashed line), and for theory including finite ion size (blue line for K$^+$ and red line for Na$^+$). b) Predictions of micropore charge density, $\sigma_{mi}$, for the theory considering point ions (dashed line) and the theory considering finite ion size (solid line). c) Predictions of the various potentials acting on finite-sized potassium ions versus $\phi_e$, including the net entropic component (black line), net electrostatic component (blue line) and net volume exclusion interactions (red line).

As the results of Fig. 2 and 3 determined that ion size plays a significant role during brackish water desalination by CDI, we now provide calculations to indicate whether CDI cells can effectively separate a wider range of ions based on their size. In Figure 4, we plot the predicted separation factor versus ion size ratio for $\phi_e = -24$ and two finite-sized and univalent counterions, where counterion "1" is Na$^+$ with $d_{Na^+} = 0.9\ nm$.

The vertical dashed lines in Fig. 4 indicate the size ratio of various cations relative to the sodium cation, which we assume to be the same size ratio as that of the hydrated cations in bulk solution (given by ref. [40]). Salt cations, such as K⁺ and lithium (Li⁺), have a size nearly equal to that of Na⁺, and thus the predicted selectivity ratios for these cations depart only modestly from unity. However, for other cations such as hydronium (H⁺) and tetrabutylammonium, larger departures are observed, with $\beta_{Na^+}/\beta_{H^+} \sim 0.5$ and $\beta_{Na^+}/\beta_{(n-Bu)_4N^+} \sim 8$.

The implication for the case of H⁺ is that volume exclusion interactions may be important to account for when attempting to model the spatiotemporal pH dynamics in a CDI cell. The value of pH within the electrodes of a charging CDI cell and also of the cell effluent is known to vary strongly relative to the feed pH,[8] which has been attributed to the effect of possible electrochemical reactions and electrosorption of H⁺ and hydroxyl (OH⁻) ions into electrode micropores.[44] Accurate modeling of the electrosorption of H⁺ and OH⁻ may require accounting for volume exclusion interactions. The calculations involving the tetrabutylammonium cation illustrates a strong effect of ion size on selectivity, as $\beta_{Na^+}/\beta_{(n-Bu)_4N^+} \sim 8$ with an ion size ratio of only ~1.4. This demonstrates that effective separation processes by ion size may be possible with a CDI cell. To date, ion separations or targeted ion removal based on ion size remains a largely unexplored application area of CDI.

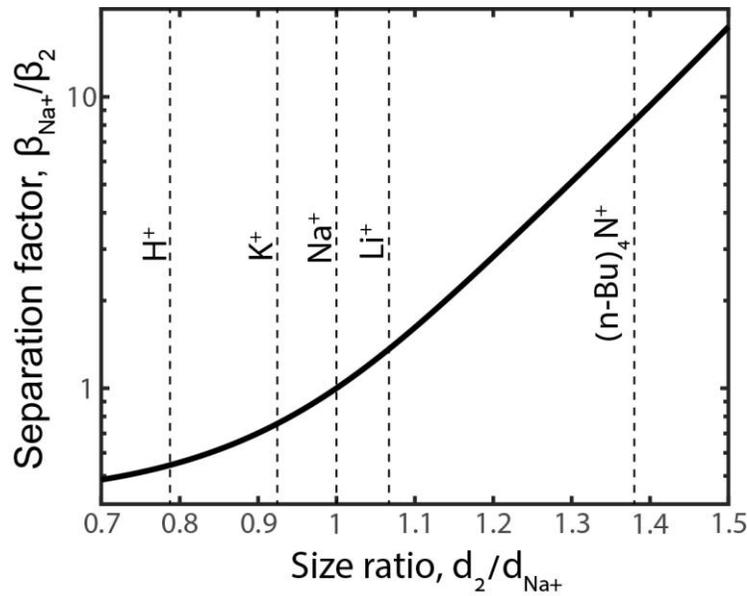

**Figure 4**: Plot of the predicted separation factor of a CDI electrode for a Na⁺ relative to a second counterion, versus the hydrated ion size ratio. Dashed lines indicate the size ratio representing various ions such as lithium and hydronium. Model parameters are $\phi_e = -24$ and $C_{st}$ = 0.2 GF/m³.

The model presented here including excluded volume interactions in micropore EDLs can be extended in several ways. Firstly, it can be generalized to the case of finite-sized ions with differing valence to capture, for example, the case of competitive electrosorption between a smaller univalent ion and a larger divalent ion. The latter scenario has been studied theoretically and experimentally in the context of charged planar interfaces,[34,35,45] and experimentally in porous electrode CDI systems,[30] but has not been incorporated into CDI theory to our knowledge. Second, while the BMCSL equation used here captures volume exclusion effects due finite-sized ions, it does not capture excess chemical potential contributions from ion-wall interactions. Such contributions may be significant in the highly geometrically-confined micropore, as suggested by previous Monte Carlo simulations showing that decreasing distance between two charged plates to approach the size of the counterions enhances the selectivity towards the smaller counterion.[33] Unfortunately, excess potentials representing wall-ion interactions are not easily described analytically, and are instead inferred from detailed molecular dynamics simulations, rendering this effect beyond the scope of this work.[46] Third, while the model presented here captures the separation factor observed in recent CDI experiments, it cannot capture other ion-size based experimental phenomena observed in microporous carbon electrodes. For example, carbon electrodes with sub-nanometer micropores have been experimentally shown to nearly completely exclude sodium ions but permit smaller hydronium ions ($\beta_{Na^+}/\beta_{H^+} \to 0$), and in other cases exhibit selectivity based on ion shape.[25,26] These phenomena may be due to strong ion-wall interactions attributed to the extremely small size of the micropores. Fourth, as our model follows a mean-field approach, it does not capture local ion-ion interactions in the micropore, which may affect the predicted micropore concentration and EDL selectivity.[38,47] Finally, the model presented here can be applied towards describing the electrosorption of large ions at higher electrode potentials than are accessible by CDI with aqueous solutions. For example, in the application of CDI to remediating organic solvents such as propylene carbonate, electrosorption of the large tetraethylammonium cations (TEA$^+$) at cell potentials above 2 V results in desolvation or solvation sheath distortion of the cation.[4] For such systems, we expect ion volume exclusion interactions to be highly important, affecting strongly the predicted equilibrium micropore concentration and the electric charge stored.

**Conclusion**

We here present a model for micropore EDLs accounting for ion volume exclusion interactions via the BMCSL equation. Our model predicts the non-unity selectivity coefficient observed experimentally in CDI systems with two counterions of equal valence, which could not be explained by previous CDI theory that assumed point ions in the micropore. Further, we show that theory based on the BMCSL equation can approximately capture the measured values of the separation factor when using the hard-sphere diameter as an adjustable parameter. In the future, this model can be extended, such as by including ion-wall or ion-ion interaction effects.

**Acknowledgments**

This work was performed in the context of a project funded by the Israeli Ministry of National Infrastructures, Energy and Water Resources, and also a project funded by the Israel Science Foundation (ISF) in the framework of the Israel National Research Center for Electrochemical Propulsion (INREP) project. We further would like to thank Maarten Biesheuvel for insightful discussions during preparation of this manuscript.